\documentclass[letterpaper]{article}

\usepackage[T1]{fontenc}

\usepackage{geometry}
\usepackage{setspace}
\usepackage{physics}
\usepackage{comment}

\usepackage{achemso}
\setkeys{acs}{articletitle = true} 

\usepackage{graphicx}
\usepackage{float}
\newfloat{scheme}{htbp}{los}
\floatname{scheme}{Scheme}
\floatname{chart}{Chart}
\newfloat{graph}{htbp}{loh}

\usepackage{chemformula} 
\usepackage[version = 4]{mhchem} 

\usepackage{authblk}
\author[1]{Pramodt Srinivasula*}
\affil[1]{Electrosoft labs LLP, Mumbai, India}
\author[2]{Rameez Raja Khan}
\author[2]{Diwakar Singh}
\author[2]{Gaurav Bhutani**}
\affil[2]{School of Mechanical and Materials Engineering, IIT Mandi, Himachal Pradesh, India}

\title{Transient Electrical Response Beyond Quasistatic Capacitance at Mechanically Excited Droplet--Dielectric Interfaces} 

\date{Email: *pramodt.research@gmail.com, **gaurav@iitmandi.ac.in}

\begin{document}

\maketitle

\begin{abstract}
Dynamic electrowetting of conducting droplets under mechanical deformation is conventionally modeled as a quasi-static variable-capacitance system, in which the electrical response is assumed to be governed solely by the evolution of the droplet--electrode contact area. Under the assumption of instantaneous charge equilibration, this framework successfully describes the cyclic steady-state electromechanical response of the system. However, its validity for transient interfacial electrical dynamics remains largely unexplored.
Here, the transient electrowetting response of mercury droplets confined between a polymeric dielectric-coated electrode (PTFE or PVDF) and an opposing copper electrode is investigated under periodic mechanical excitation with multiple waveforms at 2 Hz. The measured contact area and the corresponding capacitance evolve nearly sinusoidally and agree closely with predictions from instantaneous surface-energy minimization, confirming that the liquid-interface mechanics remain quasi-static. Likewise, the predicted scaling of cycle maximum power agrees well with experiments. In contrast, the measured transient current and instantaneous electrical power exhibit pronounced asymmetric excitation and relaxation phases that are independent of the excitation waveform, demonstrating that transient charge evolution cannot be inferred from the instantaneous geometric capacitance alone. 
This transient behavior is phenomenologically interpreted using constituent first-order interfacial dielectric charge-relaxation kinetics, indicating that the measured current arises from slow dielectric charging followed by dielectric relaxation over the timescale of the imposed periodic mechanical oscillations during discharging.
These findings establish that transient electrowetting is governed by the coupled interplay of droplet electrohydrodynamics and dielectric interfacial polarization, requiring a constitutive description beyond quasi-static variable-capacitance models based solely on contact-line dynamics.
\end{abstract}

\section*{Keywords}

Transient electrowetting; Interfacial electrical memory; Charge relaxation dynamics; Dielectric polarization; Contact capacitance.

\section{1. Introduction}

Electrowetting is an electrocapillary phenomenon in which an applied electric field modifies the apparent wettability of a conductive liquid on a dielectric-coated electrode through the coupled interactions of liquid--solid interfaces, dielectric layers, electric fields, and interfacial charge \cite{sedev2011electrowetting}. It is broadly classified into static electrowetting, where the equilibrium contact angle is electrically controlled, and dynamic electrowetting, where time-dependent wetting, spreading, and contact-line motion govern transient interfacial behavior \cite{chen2014electrowetting}. These principles underpin numerous technologies, including digital microfluidics, programmable droplet transport, adaptive optics, reflective displays, tunable liquid lenses, electrostatic actuation, and, more recently, reverse electrowetting-based mechanical energy harvesting \cite{mugele2005electrowetting,pollack2002electrowetting,li2020current,hennig2024actuating,kim2021droplet}.

Reverse electrowetting-on-dielectric (REWOD) exploits the same electrocapillary coupling in the opposite direction, whereby mechanical deformation of a liquid bridge or droplet produces electrical energy through a time-varying electrode--electrolyte capacitance \cite{verheijen1999reversible,verplanck2007reversible}. Instead of using an electric field to modulate wetting, external mechanical excitation cyclically changes the liquid geometry and interfacial contact area, inducing charge redistribution and generating alternating voltage and/or current \cite{basset2009batch,krupenkin2011reverse,yang2021laminated}. Modern REWOD devices have evolved into two distinct configurations. In electret-based REWOD \cite{adhikari2022advancing}, a permanently charged dielectric provides the internal electric field, eliminating the need for an external bias while enabling direct generation of AC voltage and current under periodic mechanical actuation \cite{kakaraparty2024advanced}. Recent studies have demonstrated self-biased operation using fluoropolymer electrets, with the generated electrical waveforms closely following the applied mechanical excitation, highlighting the strong coupling between interfacial deformation, capacitance modulation, and electrical output \cite{kakaraparty2024advanced,tao2020origami}.

Another widely investigated REWOD configuration employs a passive dielectric layer in conjunction with an externally applied DC bias \cite{yang2017high,yang2021laminated}. In these systems, the bias voltage establishes the electrostatic energy stored across the dielectric, while periodic mechanical excitation continuously modulates the liquid--electrode contact area and hence the interfacial capacitance, resulting in periodic charge transfer and electrical power generation. Owing to their simplicity, lower-cost and higher energy output, biased REWOD devices have been a dominant platform for investigating electrowetting-based energy harvesting \cite{yang2021laminated,singh2022energy}. Consequently, previous studies have primarily evaluated device performance through cycle-averaged quantities such as RMS voltage, average current, harvested power, power density, and energy-conversion efficiency. In contrast, the transient evolution of the generated current during individual wetting and dewetting cycles, which directly reflects the underlying interfacial charge-transfer dynamics, has received comparatively little attention.

The current understanding of REWOD is founded on coupled capillary and electrostatic descriptions of dynamic wetting \cite{sedev2011electrowetting}. Droplet deformation is commonly described through the Young--Lippmann relation together with surface-energy minimization to determine the evolving contact angle and liquid--electrode contact area \cite{bahadur2007electrowetting,mchale2019apparent,kim2021droplet}. The resulting capacitance variation is then coupled to electrostatic models, typically represented as a time-dependent capacitor, while equivalent RC circuit models are frequently employed to predict the generated voltage, current, and harvested power. These frameworks have successfully captured the macroscopic electrical response of REWOD devices \cite{adhikari2021electrode,tasneem2022self}.
These descriptions implicitly rely on a quasi-static approximation, wherein interfacial charge redistribution is assumed to occur much faster than droplet deformation. For conducting droplets, electrical double-layer charging typically relaxes on timescales far shorter than the mechanical excitation frequencies commonly employed in REWOD ($\sim 1$ Hz-- $1$ kHz). Consequently, the interfacial capacitance is treated as being in instantaneous equilibrium, with its temporal variation determined solely by the evolving liquid--electrode contact area.

This quasi-static picture, however, neglects a second class of interfacial phenomena associated with the dielectric layer itself. Independent studies on dielectric interfaces have shown that electrical response is often governed by finite-rate processes \cite{quader2024dielectric}, including charge trapping and detrapping \cite{verheijen1999reversible,teyssedre2021charge}, leakage conduction, Maxwell--Wagner interfacial polarization \cite{huan2026multi}, and dielectric absorption \cite{molinie2023dielectric}. Unlike the nearly instantaneous atomic and molecular polarization, these mechanisms involve charge migration and accumulation over mesoscopic length scales, exhibiting comparatively rapid polarization but much slower relaxation \cite{hamza2011relaxation}. Consequently, the dielectric interface can retain a memory of its previous electrical state \cite{chai2025high}, giving rise to hysteretic and history-dependent behavior during dynamic electrical operation \cite{ghasemi2021polymeric}. 

Dielectric interfacial charge dynamics are a central driver of dynamic electrowetting because wetting changes depend not just on capacitance, but on how charge is injected, trapped, relaxed, convected, and redistributed near the contact line and across dielectric interfaces. 
Across the literature, some of the clearest reported dynamic consequences are contact-angle saturation, hysteresis, polarity and frequency dependence, and transient spreading instabilities beyond a purely quasi-static Young--Lippmann response \cite{reid2020stick,li2017frequency}. While more emphasis of such works has been on the hydrodynamic aspects of the interfacial interactions, the role of dielectric interfacial charge dynamics in dynamic electrowetting electrical response is still limited.
Consequently, constitutive descriptions and experimental evidence that resolve the interplay between periodic mechanical forcing and dielectric interfacial charge dynamics beyond the geometrically varied capacitance limit remain lacking. In particular, transient electrical current and power responses that isolate the contribution of dielectric interfacial phenomena under periodic mechanical excitation with a fixed DC bias have not been systematically reported.

In this work, transient experiments are conducted to investigate the dynamic electromechanical response of REWOD systems, interpreted phenomenologically through a constitutive first-order kinetic model for nonequilibrium dielectric response. It is shown that although the mechanical deformation directly governs the droplet geometric-dependent capacitance variations, the transient electrical response is controlled by finite dielectric polarization and charge-relaxation dynamics beyond the conventional quasi-static description.

\section{2. Methodology}
\subsubsection{2.1 Experimental methodology}
\begin{figure}
    \centering
    \includegraphics[width=0.75\linewidth]{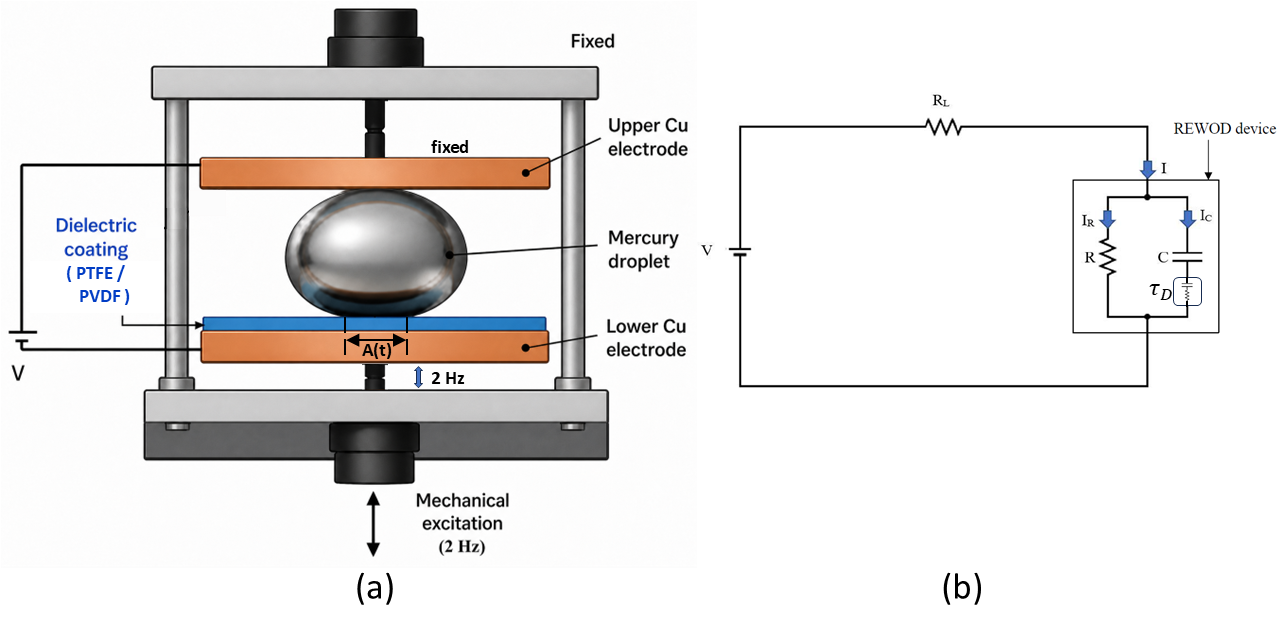}
    \caption{\textit{Experimental configuration and equivalent electrical model of the REWOD device.} (a) Illustration of the reverse electrowetting-on-dielectric (REWOD) experimental setup consisting of a mercury droplet confined between two parallel Copper electrodes, while the lower surface is coated with dielectric layers (PTFE or PVDF). Periodic mechanical excitation varies the electrode separation, causing reversible deformation of the droplet and a corresponding change in the wetted contact area, $A(t)$, and device capacitance. The generated electrical response is measured under an applied DC bias. (b) Equivalent electrical circuit of the REWOD device, modeled as a variable capacitor $C(t)$ in parallel with the dielectric leakage resistance $R$, connected to an external load resistance $R_{\mathrm{L}}$. The total current $I$ divides into the leakage current $I_R$ through the dielectric and the capacitive charging current $I_C$. An additional dielectric contribution resulting in a charge dynamics timescale $\tau_D$ in included for the transient dielectric charge-relaxation model developed in this work.}
    \label{fig:1}
\end{figure}
The REWOD experiments were performed using the parallel-plate configuration illustrated in Fig.~\ref{fig:1}(a). The device consisted of a metal--insulator--metal structure comprising a dielectric-coated copper bottom electrode, a copper top electrode, and a $25~\mu$L mercury droplet confined between the two plates. The lower electrode was subjected to controlled normal vibrations using an electrodynamic shaker while a constant DC bias voltage was applied across the REWOD cell through an external load resistance. The periodic modulation of the liquid--dielectric contact area during compression and relaxation of the droplet produced cyclic variations in capacitance, generating alternating charging and discharging currents in the external circuit.

PTFE and PVDF were selected as representative fluoropolymer dielectrics owing to their differences in surface wettability and dielectric properties. PTFE was deposited from a commercial 60 wt.\% aqueous dispersion (supplied by \textit{Sigma-Aldrich}) by spin coating at 4000 rpm for 45 s followed by curing at $120^\circ$C for 20 min. PVDF films were prepared from a 15 wt.\% solution in dimethylformamide (DMF), spin coated at 2500 rpm for 45 s and annealed at $100^\circ$C for 30 min. 

The fabricated dielectric coatings were characterized using scanning electron microscopy (SEM) to examine their surface morphology and cross-sectional thickness. Spectroscopic ellipsometry determined a dielectric thickness of $4~\mu$m for both PTFE and PVDF coatings, while fitting with the Cauchy--Urbach model yielded relative permittivities of 2.25 and 10, respectively. Static mercury contact angles, measured using a drop shape analyzer, were $154.8^\circ$ for PTFE and $141.5^\circ$ for PVDF, confirming the distinct wettability of the two dielectric surfaces. The specific capacitance of the dielectric layers was characterized by C--V measurements using a \textit{Keithley} semiconductor characterization system equipped with a micro-probing station, after thermally depositing an aluminum top electrode over an area of $270~\mu\mathrm{m}^2$. The measured specific capacitances were $6.64~\mu\mathrm{F}/\mathrm{m}^2$ for PTFE and $29.5~\mu\mathrm{F}/\mathrm{m}^2$ for PVDF.

During REWOD operation, the voltage across the external load resistance was recorded using a digital oscilloscope (manufactured by \textit{Keysight}), from which the instantaneous current was obtained using Ohm's law. Simultaneously, droplet dynamics were captured using a \textit{Photron Fastcam SA 1.1} high-speed camera operating at 1000 fps using an LED light source, enabling measurement of the instantaneous contact area and contact angle throughout the vibration cycle. The experimentally measured current waveforms together with the image-derived contact area were subsequently used to determine the instantaneous capacitance, harvested power, and energy density, discussed later.

\subsection{2.2 Theoretical framework}
\subsubsection{2.2.a Quasi-static droplet shape evolution and peak electrical power}

The experimentally observed droplet deformation during REWOD was modelled as a sequence of quasi-static equilibrium configurations using the \textit{Surface Evolver} open source code \cite{brakke1992surface}. Owing to the low excitation frequencies employed in the experiments ($2$ Hz), the characteristic capillary relaxation time of the mercury droplet is much smaller than the mechanical forcing period. Consequently, the droplet was assumed to attain equilibrium at every instantaneous plate separation. The imposed inputs to the model were the droplet volume, plate separation, applied bias voltage, dielectric thickness and dielectric constant, while the equilibrium droplet shape was obtained by minimizing the electrocapillary free energy. Electrowetting was incorporated through the Young--Lippmann equation, which relates the voltage ($V$)-dependent quasi-static equilibrium contact angle ($\theta (t)$) to the dielectric properties,

\begin{equation}
\cos\theta(V)=\cos\theta_Y+\frac{\varepsilon_0\varepsilon_rV^2}{2\gamma d},
\label{eq:lippmann_young}
\end{equation}
where $\theta_Y$ is the Young contact angle, $\gamma$ is the liquid--air surface tension, $\varepsilon_0,\varepsilon_r$ are the permittivity of free space and uniform relative permittivity of the dielectric layer, and $d$ is uniform dielectric thickness. Thus, the applied voltage modifies the solid--liquid interfacial energy through the equilibrium contact angle rather than acting directly as a body force.

For a prescribed plate separation and applied voltage, the equilibrium droplet configuration was obtained by minimizing the total interfacial free energy, comprising the liquid--air, solid--liquid, and solid--air interfacial areas, $A_{LG}$, $A_{SL}$, and $A_{SG}$, with corresponding interfacial tensions $\gamma$, $\gamma_{SL}(V)$, and $\gamma_{SG}$, respectively,
\begin{equation}
E=\gamma A_{LG}+\gamma_{SL}(V)A_{SL}+\gamma_{SG}A_{SG},
\end{equation}
subject to conservation of the droplet volume. Since $A_{SL}+A_{SG}$ remains constant for the rigid substrate, the above expression reduces to\begin{equation}
E=\gamma\left(A_{LG}-A_{SL}\cos\theta(V)\right),
\label{eq:energy}
\end{equation}
which represents the electrocapillary energy functional minimized in the present work. Accordingly, the equilibrium interface was determined by solving the constrained variational problem
\begin{equation}
\min_{\Omega} E(\Omega),
\qquad
\text{subject to}
\qquad
\mathcal V(\Omega)=\mathcal V_0,
\end{equation}
where $\Omega$ denotes the liquid interface and $\mathcal V_0$ is the prescribed droplet volume.

The liquid interface was discretized into triangular facets whose vertex coordinates constituted the numerical degrees of freedom. \textit{Surface Evolver} iteratively displaced the mesh vertices in the direction of the negative energy gradient while enforcing the constant-volume and geometric constraints corresponding to the prescribed plate separation. Adaptive step-size control, mesh refinement and Hessian-based optimization were employed to accelerate convergence to the minimum-energy configuration. The equilibrium droplet profile, contact angle and liquid--dielectric contact area obtained from the converged solution were subsequently used to determine the instantaneous capacitance and compared directly with the high-speed images acquired experimentally at corresponding stages of the vibration cycle. The experimental droplet profiles were used exclusively for validation and were not employed as inputs to the numerical model.

The equilibrium droplet profiles obtained from the numerical model were used to determine the instantaneous liquid--dielectric contact area, $A(H,\theta)$, as a function of the plate separation $H(t)$. 
The electrical circuit analogy of the present system can be simplified as a dielectric layer capacitor sandwiched between a rigid electrode (copper tape) and a deformable conductor (mercury droplet) that periodically varies the contact area with the dielectric. Fig \ref{fig:1}(b) elaborates the electrical circuit analogy.
Hence, the overall REWOD device was approximated as a parallel-plate capacitor with capacitance
\begin{equation}
C(t)=\frac{\varepsilon_0\varepsilon_rA(H,\theta)}{d},
\end{equation}
where $\varepsilon_0$ is the vacuum permittivity. Under an applied DC bias voltage $V_b$, the instantaneous equilibrium charge stored at the liquid--dielectric interface is given by $Q_{eq}(t)=V_bC(t)$, where $t$ denotes time. Consequently, the current generated during periodic droplet deformation $I_{eq}(t)=\frac{dQ_{eq}}{dt}$ was evaluated as
\begin{equation}
I_{eq}(t)=K\frac{dA}{dt}, \qquad \text{for }K=\frac{\varepsilon_0\varepsilon_rV_b}{d}.
\end{equation}
The instantaneous electrical power delivered to an external load resistance $R_L$ was calculated as $P_{eq}(t)=V_b I_{eq}(t)$, and the harvested electrical energy over one vibration cycle of period $T_m$ was obtained from $E_{eq}=\int_0^{T_m} \!\!P_{eq}(t)\,dt$.
The instantaneous maximum power was determined using the peak cyclic value of $P_{eq}$.  The power and energy density were subsequently determined by normalizing them by the maximum wetted contact area, $A_{\max}$, predicted by the numerical model. 



\subsubsection{2.2.b Phenomenological model for dielectric interfacial charge dynamics}

The classical REWOD model assumes that the electrical double layer responds instantaneously to the mechanically induced variation in the liquid--dielectric contact area, such that the interfacial charge follows the instantaneous geometric capacitance. In practice, however, the dielectric layer exhibits a finite electrical response associated with polarization, charge trapping, dielectric absorption, and related interfacial processes. Consequently, the interfacial charge cannot adjust instantaneously to changes in the equilibrium capacitance, but instead evolves over a characteristic dielectric response time.

To account for this finite response, the actual interfacial charge, $Q(t)$, is assumed to relax (either during growth or decay phase) towards its instantaneous equilibrium value, $Q_{\mathrm{eq}}(t)$, through a first-order kinetic process,
\begin{equation}
\frac{dQ}{dt}
=
-\frac{Q-Q_{\mathrm{eq}}(t)}{\tau_D},
\label{eq:charge_relaxation}
\end{equation}
where $\tau_D$ denotes the characteristic dielectric response time associated with polarization and interfacial charge relaxation. Since the experimentally measured current arises from the temporal evolution of the stored interfacial charge, it is obtained directly from the Eq. \ref{eq:charge_relaxation} as $I(t) \!=\!dQ/dt$. Equation~(\ref{eq:charge_relaxation}) therefore serves as the constitutive relation governing the transient electrical response of the dielectric interface, linking the mechanically varying equilibrium charge to the measured current. Here any other Ohmic leakage current through the dielectric (such as using a linear resistor model) is approximated to be small or less significant towards the qualitative transient behavior of electrical response and ignored. 

The proposed first-order kinetic model admits a general form of analytical solution for an arbitrary time-varying equilibrium charge arising from the applied waveform of mechanical excitation. Since the equilibrium charge $Q_{\mathrm{eq}}(t)$ follows the instantaneous capacitance based on the contact area, Eq.~(\ref{eq:charge_relaxation}) constitutes a linear first-order ordinary differential equation as follows.
\begin{equation}
Q(t)
=
Q_1e^{-(t-t_1)/\tau_D}
+
\frac{\varepsilon_0\varepsilon_rV_b}{\tau_Dd}
\int_{t_1}^{t}
A(s)e^{-(t-s)/\tau_D}\,ds,
\label{eq:Q_general}
\end{equation}
where $Q_1 \!=\! Q(t_1)$ denotes the initial interfacial charge at an arbitrary reference time $t_1$. Equation~(\ref{eq:Q_general}) shows that the transient charge is governed by a convolution of the contact-area history with an exponential relaxation kernel, indicating that the stored charge depends on both the instantaneous droplet geometry and its previous evolution due to the interfacial dielectric response.

Using Eq.~(\ref{eq:charge_relaxation}) together with Eq.~(\ref{eq:Q_general}), and integrating the convolution by parts, gives
\begin{equation}
I(t)
=\frac{K}{\tau_D}
\int_{t_1}^{t}
e^{-(t-s)/\tau_D}
\frac{dA}{ds}\,ds
+
I_1e^{-(t-t_1)/\tau_D},
\label{eq:I_general}
\end{equation}
where $I_1 \!=\! -(Q_1-Q_{\mathrm{eq},1})/\tau_D$ and $Q_{\mathrm{eq},1} \!=\! KA(t_1)$ are the current and equilibrium charge estimation at the instance of reference time $t_1$.
Equation~(\ref{eq:I_general}) represents the general hereditary integral constitutive relation for the transient electrical response of the REWOD interface. It identifies two distinct current contributions arising naturally from the proposed first-order dielectric kinetics: (i) a history-dependent hereditary contribution arising from finite-rate dielectric relaxation, and (ii) a transient contribution associated with the relaxation of the initial nonequilibrium charge at the reference state. 

The relative importance of these mechanisms is governed by the dimensionless ratio of the characteristic time scale of the dielectric relaxation and timescale of the mechanically induced contact-area variation (such as the time period of mechanical oscillation), $\alpha \!=\! \tau_D/T_m$. While the description of the system simplifies to equilibrium dynamics for $\alpha \!\ll \! 1$,  the limiting behaviors corresponding to $\alpha\sim1$, and $\alpha\gg1$ are discussed subsequently in the context of the experimental observations during the charging and discharging phases of periodic mechanical vibrations.
During the spreading phase, the reduction in electrode separation ($d$) increases both the liquid--dielectric contact area and the electric field across the dielectric, promoting dielectric polarization and interfacial charge accumulation. As the droplet subsequently retracts, the electrode separation increases, reducing the equilibrium interfacial charge and electric field. Polymeric dielectric systems are known to exhibit asymmetric charging and relaxation kinetics \citep{hamza2011relaxation,ghasemi2021polymeric}. 
While the scope of this paper excludes the microscopic details of charge dynamics in the dielectrics, the interfacial phenomena can still be captured using the phenomenological model introduced here, using two independent characteristic time scales, $\tau_D \!=\! \tau_c$ during the interfacial charging phase and $\tau_D \!=\! \tau_r$ during the discharging phase.







\section{3. Results and Analysis}

\subsubsection{3.1 Mechanical excitation oscillations }

\begin{figure}
    \centering
    \includegraphics[width=1\linewidth]{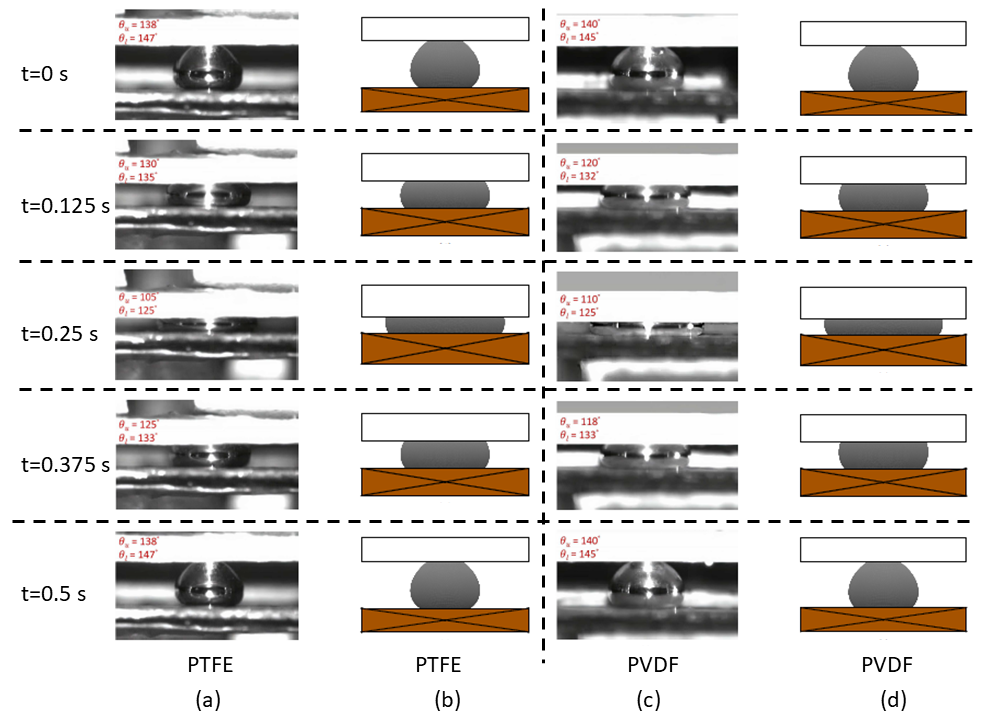}
    \caption{\textit{Transient droplet deformation during one mechanical excitation cycle.} Time-lapse images and corresponding \textit{Surface Evolver} simulations of a  $25$ $\mu$L Mercury droplet deformation under sinusoidal mechanical excitation for (a,b) PTFE and (c,d) PVDF dielectric coatings. The experimental and simulated droplet profiles show excellent agreement throughout the excitation cycle.} 
    \label{fig:2}
\end{figure}

The transient deformation of the REWOD droplet under sinusoidal mechanical excitation is first examined by comparing the experimentally observed droplet profiles with the equilibrium shapes predicted from surface energy minimization. Figures~\ref{fig:2}(a--d) present representative snapshots over one siunusoidal waveform oscillation cycle at 2 Hz for the PTFE and PVDF dielectric coatings. For both materials, the droplet undergoes a highly repeatable periodic spreading and recoiling motion synchronized with the imposed mechanical vibration. Starting from its nearly spherical configuration at $t=0$ s, the droplet progressively spreads as the electrode separation decreases, reaches its maximum deformation at approximately $t=0.25$ s, and subsequently retracts to recover its initial shape by $t=0.5$ s, completing one excitation cycle. The excellent repeatability of this deformation confirms that the droplet motion is predominantly governed by capillary equilibrium under the slowly varying mechanical boundary condition.
The measured advancing and receding contact angles indicated in the experimental images follow the same temporal trend predicted by the surface energy minimization model, demonstrating that the quasi-static equilibrium assumption remains valid under the low-frequency sinusoidal excitation employed in the present study. However, minor deviations are observed in the contact angles during the spreading and de-spreading phases in the experiments, while the simulations estimate nearly identical shapes. This is primarily attributed to contact line hysteresis which are not explicitly included in the equilibrium calculations.

\begin{figure}
    \centering
    \includegraphics[width=0.91\linewidth]{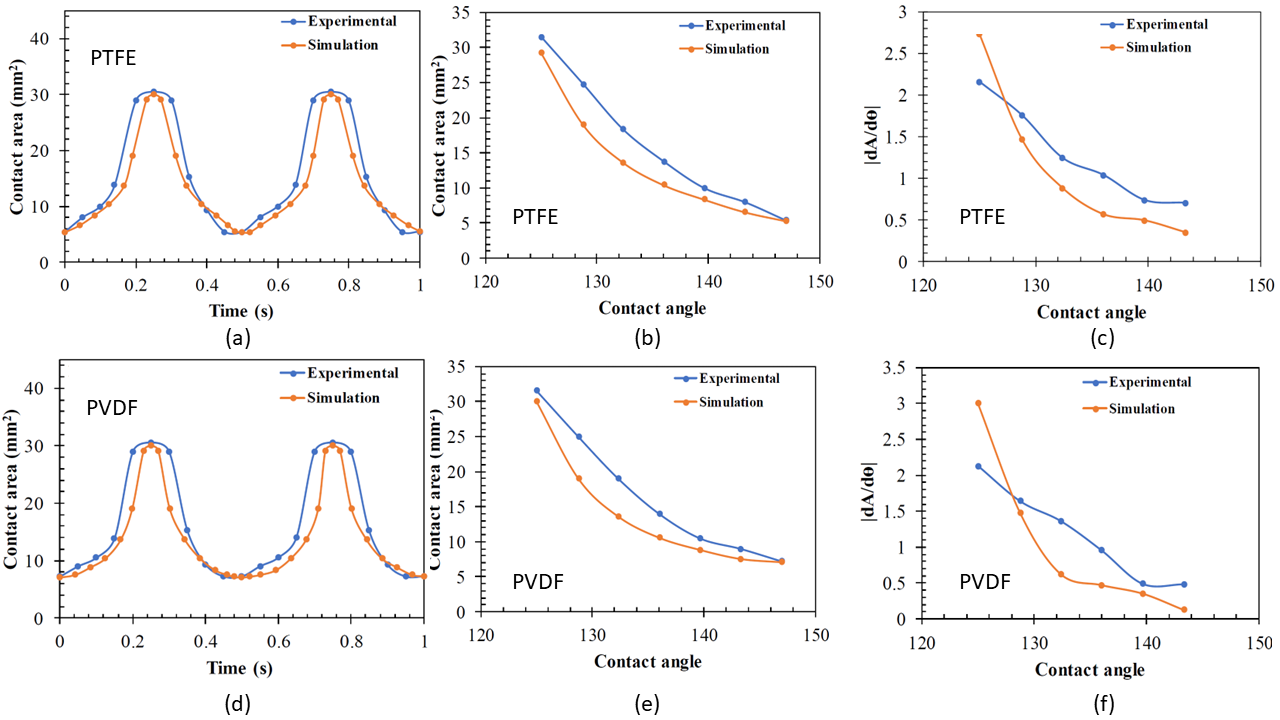}
    \caption{\textit{Correlation between contact angle and contact area during transient droplet oscillation.} Comparison of experimental measurements and surface energy minimization calculations for the PTFE and PVDF dielectric coatings for the case of Fig. \ref{fig:2}.  Temporal variation of the wetted contact area for (a) PTFE and (d) PVDF, comparing experimental measurements with surface energy minimization calculations.  Contact area as a function of contact angle for (b) PTFE and (e) PVDF, corresponding area sensitivity, $|dA/d\theta|$, for (c) PTFE and (f) PVDF}
    \label{fig:3}
\end{figure}

The temporal evolution of the contact area further quantifies this agreement between the experiments and simulations. Figures~\ref{fig:3}(a,d) compare the experimentally measured contact area with the values predicted from the \textit{Surface Evolver} calculations for the PTFE and PVDF coatings, respectively. In both cases, the contact area varies periodically following the imposed sinusoidal excitation, increasing from approximately $5$--$7~\mathrm{mm^2}$ in the relaxed state to nearly $30~\mathrm{mm^2}$ at maximum compression before returning to its initial value. The simulated curves closely reproduce both the amplitude and phase of the experimentally measured oscillations throughout successive vibration cycles, indicating that the geometrical evolution of the droplet is accurately captured by the surface energy minimization subject to mechanical confinement. The nearly identical periodic response over consecutive cycles also confirms the absence of measurable drift or degradation during repeated mechanical excitation.
Since the capacitance of the REWOD device is directly proportional to the instantaneous liquid--solid contact area, the agreement between the experimental and simulated droplet deformation establishes the validity of the geometrical model used in the subsequent electrical analysis. 

Figures~\ref{fig:3}(b,e) compare the experimentally measured contact area with the predictions obtained from the surface energy minimization model as a function of the contact angle. For both dielectric coatings, the contact area increases monotonically and nonlinearly as the contact angle decreases, with excellent agreement between the experiments and simulations throughout the entire deformation range. The largest contact areas are attained at the minimum contact angles corresponding to the maximum compression of the droplet, whereas the contact area decreases continuously as the droplet relaxes towards its nearly spherical equilibrium shape. The close correspondence confirms that the transient droplet geometry is predominantly governed by capillary surface energy minimization under constant droplet volume.

Figures~\ref{fig:3}(c,f) further quantify the geometric sensitivity by plotting the local area gradient, $|dA/d\theta|$, as a function of the contact angle for PTFE and PVDF, respectively. The contact angle determines the contact-line force balance and the shape of the liquid--gas interface while conserving the total enclosed droplet volume, thereby governing the droplet spreading dynamics. Both the experimental measurements and the surface energy minimization calculations exhibit a pronounced nonlinear decrease in $|dA/d\theta|$ with increasing contact angle, indicating that the wetted area becomes progressively less sensitive to variations in the contact angle as the droplet retracts. This reduction in geometric sensitivity implies that identical changes in the contact angle produce progressively smaller changes in the contact area during the retraction phase.


\subsection{3.2 Electrical response}
\subsubsection{3.2.a Cycle-average or cycle maximum of response}
\begin{figure}
    \centering
    \includegraphics[width=0.65 \linewidth]{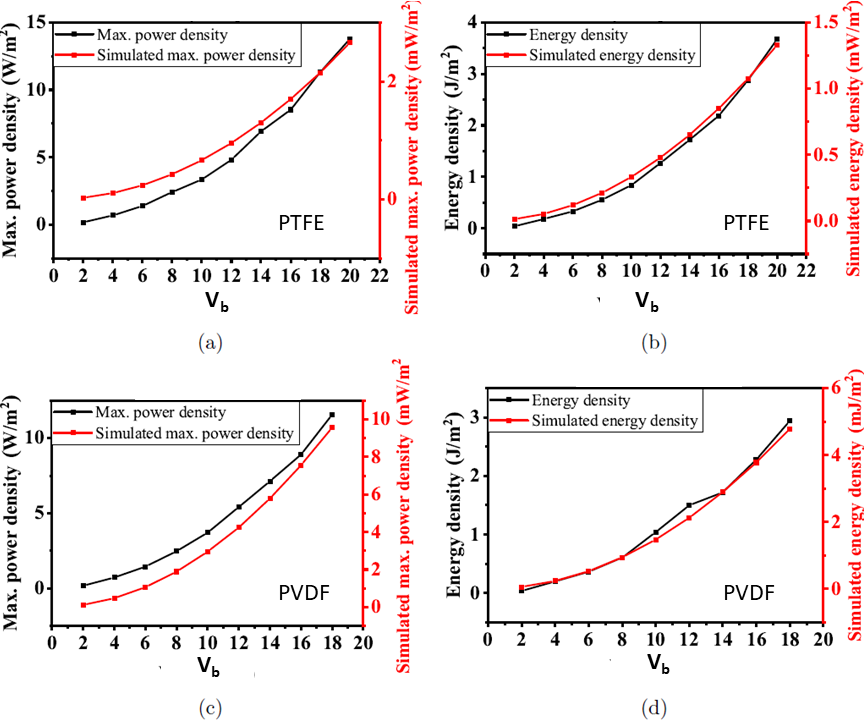}
    \caption{\textit{Voltage dependence of the electrical power and energy density.} Comparison of experimental measurements and model predictions at a mechanical excitation frequency of 2 Hz for PTFE and PVDF dielectric coatings. Maximum instantaneous power density as a function of the applied DC bias voltage for (a) PTFE and (c) PVDF, and corresponding harvested cycle-averaged energy density for (b) PTFE and (d) PVDF.}
    \label{fig:4}
\end{figure}
Figure~\ref{fig:4} presents the influence of the applied DC bias voltage on the maximum instantaneous power density and cycle-averaged harvested energy density for PTFE and PVDF dielectric coatings at a constant excitation frequency of 2 Hz. Figures~\ref{fig:4}(a,c) compare the experimentally measured maximum power density with the estimates based on quasi-static simulations. For both dielectrics, the maximum power density increases nonlinearly with the applied bias voltage, in excellent qualitative agreement with the simulations. This behavior is consistent with the quadratic scaling of electrostatic energy with voltage in electrowetting, $P\sim V_b^2$ inline with quasistatic estimates based on geometric capacitance variation. The corresponding harvested energy densities are shown in Figs.~\ref{fig:4}(b,d), increasing nonlinearly with the applied bias voltage, with the simulations accurately reproducing the experimental measurements.

\subsubsection{3.2.b Transient response}
\begin{figure}
    \centering
    \includegraphics[width=0.7\linewidth]{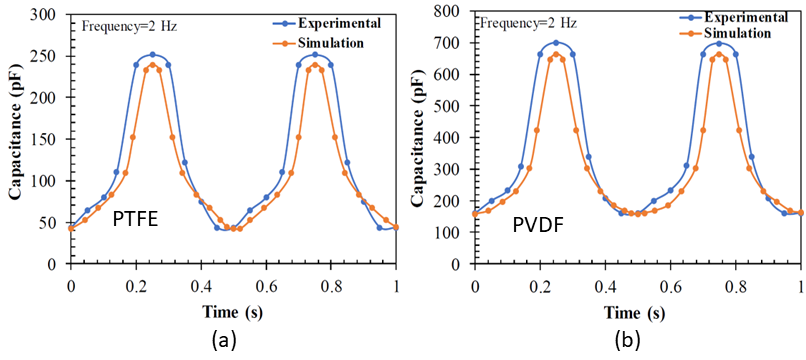}
    \caption{\textit{Capacitance temporal variation  for the case of Fig. \ref{fig:2}} with (a) PTFE- (b) PVDF-dielectric coating.}
    \label{fig:5A}
\end{figure}
\begin{figure}
    \centering
    \includegraphics[width=0.65\linewidth]{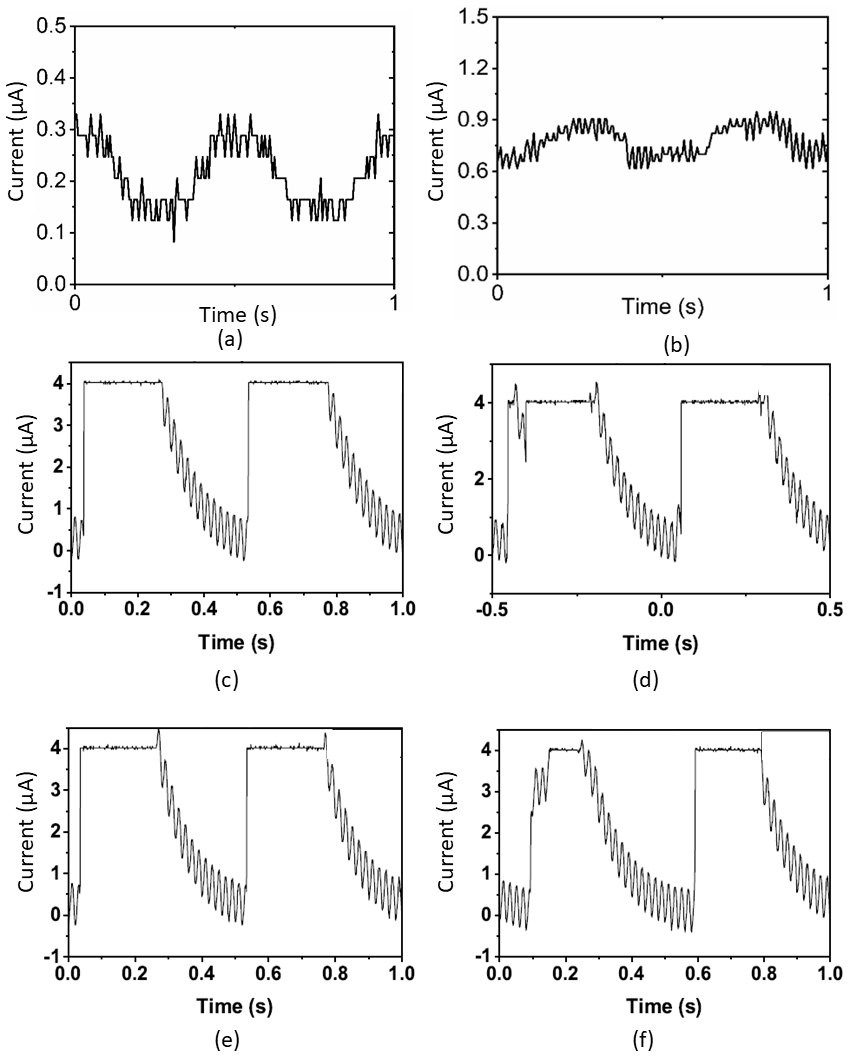}
    \caption{\textit{Transient power response under different mechanical excitation waveforms.} Experimentally measured electrical current for REWOD devices at 4 V bias voltage with (a) TiO$_2$, (b) ZnO+PTFE composite, and (c--f) PTFE dielectric coatings under a 2 Hz mechanical excitation. The applied excitation waveforms are (a--c) sinusoidal, (d) square, (e) triangular, and (f) ramp. The ceramic and composite dielectric coatings respond instantaneously to mechanical excitation in (a,b), whereas the polymer-based PTFE coating exhibits a waveform-independent two-timescale response consisting of rapid excitation followed by slow dielectric relaxation in (c--f).}
    \label{fig:5}
\end{figure}

The resulting temporal capacitance variation is presented in Figs.~\ref{fig:5A}(a,b) for the PTFE and PVDF coatings, respectively. Since the instantaneous capacitance is directly proportional to the wetted contact area, the capacitance follows the same periodic evolution as the droplet deformation, exhibiting maxima corresponding to the instants of maximum spreading. The capacitance predicted from the simulated contact area closely reproduces both the amplitude and phase of the experimentally measured capacitance. As expected from its higher dielectric constant, the PVDF coating produces substantially larger capacitance values than PTFE while maintaining a similar temporal evolution.

Similarly, under instantaneous charge dynamics, a sinusoidal mechanical excitation is expected to produce a sinusoidal transient electrical current. This expectation is consistent with the behavior observed for ceramic and composite dielectric coatings. Figures~\ref{fig:5}(a,b) show the transient current measured from TiO$_2$ and ZnO nanowire/PTFE composite dielectric coatings, respectively, both exhibiting nearly sinusoidal responses that closely follow the mechanical excitation. Additional details of the materials, fabrication, and experimental procedures are provided in Appendix~A.

In contrast, the polymer-based PTFE dielectric exhibits a qualitatively different transient electrical response. Figures~\ref{fig:5}(c--f) present the current generated by the PTFE-coated REWOD device under a 2 Hz mechanical excitation at a bias voltage of $V_b=4$ V for four excitation waveforms: sinusoidal, square, triangular, and ramp. In every case, the electrical current rises rapidly to a nearly constant value immediately after the onset of mechanical excitation during the charging phase, as the electrodes move closer together and the droplet spreads over the dielectric surface. This is followed by a much slower decay over a time scale of $0.25$ s toward nearly zero during the discharging phase, which occupies the remaining half of the oscillation cycle before the next charging event begins. This asymmetric transient response is observed irrespective of the excitation waveform and the bias voltage up to dielectric breakdown. Very similar results were observed with PVDF coating as well (although not shown here). The pronounced contrast between the nearly sinusoidal variations in the mechanical displacement and capacitance, and the strongly asymmetric electrical current, demonstrates that the electrical output cannot be explained solely by the instantaneous capacitance variation of an ideal variable capacitor. Instead, the response is characterized by the more complex transient interfacial dielectric phenomena of the polymeric materials.

The phenomenological model developed in Section~2.2.b expresses the transient current as the superposition of two constitutive contributions arising from the history-dependent dielectric response, and the relaxation of the initial nonequilibrium electrical state. As summarized in Appendix~B, the relative importance of these contributions in Eq. \ref{eq:I_general} depends on the dimensionless dielectric response parameter $\alpha$, giving rise to three distinct constitutive regimes corresponding to fast ($\alpha\ll1$), intermediate ($\alpha\sim1$), and slow ($\alpha\gg1$) dielectric relaxation.

The experimentally observed current waveforms can be interpreted within this framework. 
During the charging phase, the current remains nearly constant despite the continuing increase in contact area. This behavior is consistent with the slow dielectric-response limit ($\alpha_c \gg 1$), in which the electrical response is only weakly modulated by the instantaneous mechanical deformation and is instead governed primarily by the initial equilibrium state. The phenomenological model therefore provides a natural interpretation for the sustained current plateau observed experimentally during charging. 
In contrast, the discharging phase exhibits the transient current in line with the case of the characteristic charge relaxation time $\tau_D\!=\!\tau_r$ comparable to the mechanical period $T_m$, i.e., $\alpha_r\sim1$, indicating that dielectric relaxation contributes significantly to the transient current evolution. Although the present phenomenological model does not separately quantify the individual constitutive contributions during each phase, it provides a physically consistent framework for interpreting the experimentally observed transition from an approximately constant charging current to a slowly relaxing discharging current.

\section{4. Discussion}

\begin{figure}
    \centering
    \includegraphics[width=1\linewidth]{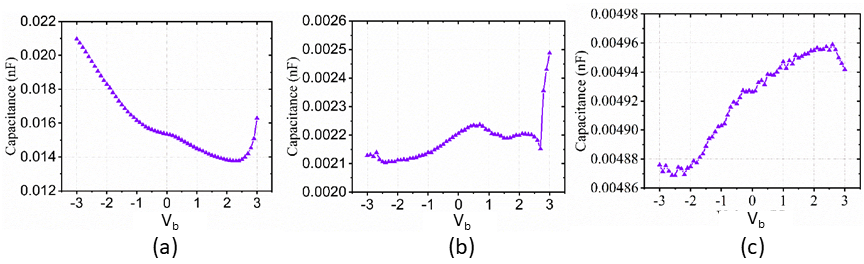}
    \caption{\textit{Voltage-dependent capacitance of different dielectric coatings.} Measured capacitance as a function of the applied DC bias voltage for dielectric layers of (a) ZnO (b) ZnO/PTFE composite, and (c) PTFE.} 
    \label{fig:8}
\end{figure}

The primary contribution of this work is the experimental identification of transient electrical response regimes in reverse electrowetting on polymeric dielectric coatings under periodic mechanical excitation. These previously unrecognized charging and discharging characteristics reveal that the underlying energy-conversion process involves coupled dielectric relaxation and electrohydrodynamic interfacial phenomena that extend beyond the conventional quasi-equilibrium electrowetting picture.

The proposed phenomenological transient dielectric model (Eq. \ref{eq:charge_relaxation}) employed in the present work represents the effective polarization and charge-relaxation dynamics through the characteristic charging and discharging timescales, without explicitly resolving the individual microscopic transport mechanisms within the dielectric layer. The model captures the existence and persistence of the charging current plateau, whereas the detailed dynamics of the transition into this plateau at the onset of the chaging phase of the periodic excitations remain beyond the scope of the present phenomenological description. 
The model also captures the qualitative exponential relaxation observed during the discharging phase, although the individual constitutive contributions to this behavior cannot be uniquely separated within the present phenomenological framework.
Accordingly, the present phenomenological model should be viewed as providing a constitutive framework for interpreting the experimentally observed charging and discharging characteristics rather than a quantitative decomposition of the individual mechanisms contributing to each transient feature. 

In practical dielectric materials, several concurrent processes may contribute to the measured electrical response, including dielectric absorption, charge trapping and detrapping, leakage conduction through localized defect states, Maxwell--Wagner interfacial polarization, and finite ionic transport along the solid--liquid interface. Their relative contributions depend strongly on factors such as the dielectric composition, microstructure, defect density, moisture adsorption, and applied electric field. Consequently, the effective relaxation time identified in the present model should be interpreted as a macroscopic constitutive parameter representing the combined influence of these interfacial dielectric processes rather than a single microscopic mechanism. The systematic evolution of the transient waveforms from PTFE to TiO$_2$ and ZnO+PTFE (in Fig. \ref{fig:5}) nevertheless demonstrates that the material-specific dielectric relaxation dominates the transient electrical response over the investigated operating conditions. 
While leakage current processes of conduction in nature may contribute appreciably through the resistive current, $I_R$ (Fig.~\ref{fig:1}(b)), they primarily affect the current magnitude rather than the characteristic relaxation timescale. They are therefore omitted from the present analysis, which focuses on the excitation--relaxation dynamics, however could be of interest for future investigations.


Further, beyond the polymeric dielectric coatings analyzed in this study, the voltage-dependent capacitance of the REWOD device could be strongly governed by the dielectric layer composition itself.
Figure~\ref{fig:8}(a-c) compares the measured capacitance as a function of the applied DC bias voltage for ceramic ZnO, composite ZnO/PTFE , and polymeric PTFE dielectric layers, respectively. Unlike the nearly monotonic increase observed for PTFE, and decrease observed for ZnO, the ZnO/PTFE coating exhibit distinctly different, non-monotonic voltage dependence, demonstrating that the capacitance is strongly influenced by the dielectric material and its microstructure. These differences reflect the interplay between dielectric permittivity, interfacial polarization, charge trapping, and field-dependent dielectric relaxation, indicating that the effective capacitance cannot, in general, be described solely by the geometric contact area. This highlights opportunities for future investigations at the intersection of electrohydrodynamics and interfacial dielectric materials and phenomena, with implications for both fundamental understanding and practical applications.

From the practical implementation and applications perspective, the operating voltage is ultimately constrained by both dielectric integrity and hydrodynamic stability of the electrowetting interface. For the present dielectric thicknesses, irreversible dielectric breakdown voltages $V_{b,max}$ was experimentally observed at approximately 20~V for PTFE and 18~V for PVDF, manifested by the formation of localized defects within the dielectric layer accompanied by a sudden loss of insulating behavior. 
In addition to dielectric failure, sufficiently large electric stresses may also induce hydrodynamic instabilities such as excessive contact-line distortion, contact angle saturation, partial depinning, or unstable droplet deformation. Although such instabilities were avoided in the results presented in this report, it is observed that the square-wave excitation is more susceptible to the onset of droplet hydrodynamic disintegration under strong confinement and electric fields than the other waveforms. These observations define the practical upper operating limit for the present REWOD configuration. Below these limitations, the electrical response remained repeatable over repeated oscillation cycles, indicating that the measured transient dynamics arise from reversible dielectric polarization rather than permanent material degradation. 

To facilitate comparison with previously reported REWOD energy harvesters employing different device dimensions, operating voltages, and excitation frequencies, the normalized figure of merit, $\mathrm{FoM}=P_{peak}/V_{b,max}^{2}f \overline A$, with mechanical excitation frequency $f \!=\! 1/T_m$, cyclic peak power $P_{peak}$ and an effective contact area during the cycle $\overline A$ can be employed in line with \citet{hsu2015bubbler}. The present devices yield FoM values of approximately $17.51\times10^{-3}\ \mathrm{W\,V^{-2}\,Hz^{-1}\,mm^{-2}}$ and $18.42\times10^{-3}\ \mathrm{W\,V^{-2}\,Hz^{-1}\,mm^{-2}}$  for the PTFE and PVDF coatings, respectively. These values are comparable with those reported in the existing REWOD literature  \cite{basset2009batch,krupenkin2011reverse,hsu2015bubbler} , demonstrating that the present experimental setup operate within the practical device performance range despite being primarily designed to investigate the fundamental transient electrowetting physics. Consequently, the dielectric-dependent excitation--relaxation dynamics identified in this work are directly relevant to practical REWOD energy harvesters and provide physically grounded guidelines for optimizing dielectric material selection in future device designs.

Although the present investigation is motivated by reverse electrowetting energy harvesting, the underlying dielectric-controlled transient behavior is expected to extend more generally to dynamic electrowetting systems operating under periodically varying electrical or mechanical forcing. Numerous applications, including digital microfluidics, adaptive liquid lenses, droplet actuation, vibration-driven energy harvesters, self-powered sensors, and fluidic switching devices, operate over characteristic frequencies ranging from sub-hertz to several tens of hertz, where the timescale of relaxation dielectric responses may become comparable to the actuation period. Hence, under these conditions, the common quasi-static assumption that the dielectric instantaneously follows the external forcing may no longer be strictly valid.

\section{5. Conclusions}
This work demonstrates that the transient electrical response of reverse electrowetting-on-dielectric systems is governed not only by the mechanically induced variation in capacitance, but also by the intrinsic dielectric polarization and charge-relaxation dynamics of the solid--liquid interface. Surface energy minimization model accurately predicted the transient droplet deformation, contact area, capacitance and cyclic steady-state electrical power for PTFE and PVDF coatings, while the experimentally observed transient current and power exhibited charging and discharging phases of distinct time scales and transient behaviors, that cannot be explained by a quasi-static variable-capacitor model alone. 
Although a detailed microscopic description of the dielectric relaxation mechanisms remains beyond the scope of this work, the experimentally observed transient electrical response is consistently interpreted using a first-order phenomenological charge-relaxation model. The model reveals quasi-static, history-dependent, and nonequilibrium relaxation contributions to the measured current, providing a constitutive interpretation of the observations as slow charging behavior followed by dielectric relaxation on the timescale of the periodic mechanical excitation during the interfacial discharging phase. 
The measured figures of merit are comparable to those reported in the energy harvesting literature, confirming the practical relevance of the present investigation. More broadly, these findings establish that dynamic electrowetting is governed by the coupled interplay of electrohydrodynamics and dielectric phenomena at solid--liquid interfaces, providing a framework for understanding and engineering transient electrowetting and related interfacial electrohydrodynamic systems.

\appendix

\renewcommand{\thefigure}{A\arabic{figure}}
\setcounter{figure}{0}

\section{Appendix A: Experimental characterization of dielectric surfaces}
The SEM images in Fig.~A1 reveal distinct microstructural characteristics of the dielectric coatings, which are expected to influence their interfacial electrical properties. The PTFE coating (Fig.~A1(a)) exhibits a compact and relatively homogeneous granular morphology, characteristic of a continuous fluoropolymer film with low surface energy. In contrast, the PVDF coating (Fig.~A1(b)) displays a porous, interconnected microstructure, providing a larger effective interfacial area. The ZnO nanowires (Fig.~A1(c)) consist of densely packed, high-aspect-ratio crystalline structures, while the cross-sectional image of the deposited ZnO coating (Fig.~A1(d)) confirms the formation of a continuous film of approximately $5~\mu$m thickness on the copper (Cu) substrate. The TiO$_2$ coating (Fig.~A1(e)) exhibits an agglomerated nanoparticulate morphology with a highly rough surface. These differences in surface morphology, porosity, and microstructure are expected to influence the effective dielectric response through variations in interfacial polarization, charge trapping, and dielectric relaxation, thereby contributing to the material-dependent transient electrical behavior observed in the present study.

\begin{figure}
    \centering
    \includegraphics[width=1\linewidth]{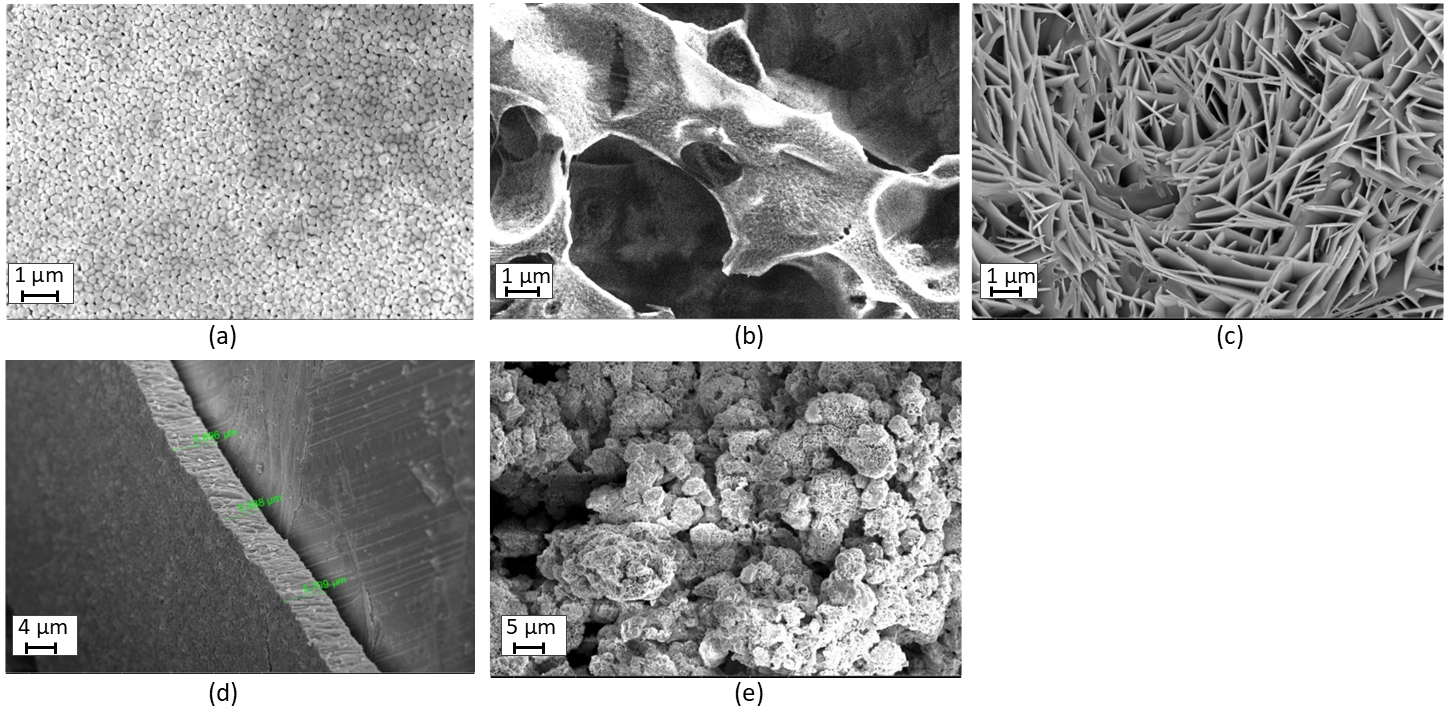}
   \caption{\textit{Scanning electron microscopy (SEM) images of the dielectric materials used in the present study.} (a) PTFE coating, (b) PVDF coating, (c) ZnO nanowire morphology prior to deposition, (d) cross-sectional SEM image of the ZnO coating deposited on Copper tape, and (e) TiO$_2$ coating deposited on Copper tape.}
    \label{fig:A1}
\end{figure}

\section*{Appendix B: Asymptotic analysis of the transient current from the phenomenological model}

\setcounter{equation}{0}
\renewcommand{\theequation}{B\arabic{equation}}

The relative importance of the two mechanistic contributions of current in Eq. \ref{eq:I_general} are analyzed below with the limiting values of $\alpha$.

\subsection*{B.1. Instantaneous dielectric response ($\alpha\ll1$)}
When the dielectric responds much faster than the mechanical deformation, the interfacial charge remains in quasi-equilibrium with the instantaneous capacitance, $Q(t)\rightarrow Q_{\mathrm{eq}}(t)=KA(t)$. The exponential kernel approaches a Dirac delta function $\frac{1}{\tau_D} e^{-(t-s)/\tau_D} \rightarrow \delta(t-s)$, while the transient term decays immediately, $I_1e^{-(t-t_1)/\tau_D}\rightarrow0$. Consequently, Eq.~(\ref{eq:I_general}) reduces to
\begin{equation}
I(t)
=
K\frac{dA}{dt},    
\end{equation}
which is the classical instantaneous variable-capacitance model employed in conventional REWOD analyses.

\subsection*{B.2. Slow dielectric response ($\alpha\gg1$)}

When the dielectric relaxation time is much larger than the mechanical period, $\tau_D\gg T_m$, the exponential kernel varies only weakly during one oscillation cycle,
\begin{equation}
  e^{-(t-s)/\tau_D}\approx1.  
\end{equation}
Hence the total current in Eq.~(\ref{eq:I_general}) reduces to 
\begin{equation}
    I(t) \!\approx\! I_1,    
\end{equation}
a time independent value, dependent only on the initial reference state nonequilibrium current. 

Physically, the equilibrium charge $Q_{eq}(t)$ continues to vary with the oscillating contact area, whereas the actual interfacial charge evolves only over the much longer timescale $\tau_D$. Consequently, the measured current becomes nearly constant over the mechanical cycle time scale and is no longer modulated by the periodic deformation. The value of this current is determined solely by the  the initial reference state contact area $A(t_1)$, independent of the applied mechanical waveform of excitation. The electromechanical coupling is therefore strongly attenuated, leaving only a weak, slowly varying (nearly constant) current associated with dielectric relaxation.

\subsection*{B.3. Comparable dielectric and mechanical timescales ($\alpha\sim1$)}

When the dielectric response time is comparable to the mechanical excitation period, neither the quasi-equilibrium approximation nor the frozen-charge approximation is valid. Both the terms in Eq.~(\ref{eq:I_general}) contribute simultaneously to the transient current. The first term, the hereditary integral accounts for the finite-rate dielectric response to the evolving contact area, and the exponential term describes relaxation from the initial nonequilibrium state. The resulting current therefore exhibits a coupled electromechanical response characterized by temporal lag, waveform distortion, and hysteretic behavior or a finite electrical memory.

The resulting current can be expressed as
\begin{equation}
I(t)
=
I_1 e^{-(t-t_1)/\tau_D}
\left[
1+
\left(\frac{\alpha_Q}{\alpha}\right)
e^{(t-t_1)/\tau_D}
\int_{t_1}^{t}
e^{-(t-s)/\tau_D}
\frac{d\hat{A}}{ds}\,ds
\right],
\label{eq:current_general}
\end{equation}

where $\hat{A} \!=\!A/\overline A \! \sim\mathcal{O}(1)$ denotes the normalized contact area with respect to a nominal area $\overline A$. The dimensionless parameter $\alpha_Q \!=\! Q_n/(I_1T_m)$, compares the nominal capacitive charge accumulated over one mechanical oscillation with the initial reference current. Here, $Q_n \!=\!\varepsilon_0\varepsilon_r \overline{A} V_b/d$ is the characteristic interfacial charge associated with the nominal contact area. Consequently, the measured current is governed by the competition between the intrinsic dielectric relaxation and the mechanically driven evolution of the liquid--solid contact area.

During the initial stage of droplet dewetting from a highly stretched configuration, the reference current $I_1$ is typically much larger than the characteristic current associated with the subsequent cyclic oscillation. Under these conditions, $\alpha_Q\ll1$. Consequently the hereditary contribution becomes asymptotically small. Equation~(\ref{eq:current_general}) therefore reduces to
\begin{equation}
    I(t)\approx
I_1e^{-(t-t_1)/\tau_r},    
\end{equation}
indicating that the current is primarily governed by the free dielectric relaxation. As the relaxation proceeds and the exponential term decays, the mechanically induced hereditary contribution could progressively becomes significant, eventually governing the periodic current response during the subsequent oscillatory motion.
\\

\textbf{Declarations}\\
\textbf{Authors' contributions: } P.S. conceived the study, analyzed the data, and wrote the original manuscript. R.R. and D.S. developed the experimental methodology, performed the experiments, and generated the experimental data. R.R. conducted the simulations. G.B. conceived the study, supervised the project, provided resources, and reviewed the manuscript. All authors contributed to the interpretation of the results.\\
\textbf{Funding acknowledgments:}  This project did not receive any external funding.\\
\textbf{Data availability:} Data will be made available on an external online sharing platform. \\
\textbf{Conflicts of interest:} Authors declare no conflicts of interest.

\bibliography{references.bib}

\end{document}